\documentclass{article}
\usepackage{a4wide}
\usepackage[english]{babel}
\usepackage{amsmath,amsfonts,amssymb,amsthm}
\newcommand{\cK}{\mathcal{K}}

\newtheorem{theorem}{Theorem}
\newtheorem{proposition}[theorem]{Proposition}
\newtheorem{corollary}[theorem]{Corollary}

\newtheorem{lemma}[theorem]{Lemma}
\numberwithin{equation}{section}

\newcommand{\by}{\mathbf{y}}
\newcommand{\bx}{\mathbf{x}}

\newcommand{\bR}{\mathbb{R}}
\newcommand{\bZ}{\mathbb{Z}}

\newcommand{\bP}{\mathbf {P}}

\newcommand{\px}{\bx_{\perp}}
\newcommand{\py}{\by_{\perp}}
\newcommand{\parX}{\langle X_{\parallel, a}\rangle}
\newcommand{\pargX}{\langle X_{\parallel, g}\rangle}
\newcommand{\perX}{\langle X_{\perp}\rangle}
\newcommand{\paralpha}{\alpha_{\parallel}}
\newcommand{\peralpha}{\alpha_{\perp}}
\newcommand{\pg}{g_{\perp}}
\newcommand{\hX}{\hat{X}_{\parallel}}
\newcommand{\Tn}{T_n^{\theta_0}}

\DeclareMathOperator{\dist}{dist}

\newcommand{\cP}{\mathcal{P}}
\newcommand{\cA}{\mathcal{A}}

\newcommand{\cref}[1]{Corollary~{\ref{#1}}}

\begin{document}

\title{ Optimally localized Wannier functions for
  quasi one-dimensional nonperiodic insulators}
\author{ H. D. Cornean$^{1}$, A. Nenciu$^{2}$, G. Nenciu$^{3}$\\
$^{1}$Department of Mathematical Sciences, Aalborg University,
 Fredrik Bajers Vej\\
7G, DK-9220 Aalborg, Denmark\\
 $^2$Faculty of Applied Sciences
University ``Politehnica'' of Bucharest,\\ Splaiul Independentei
313, RO-060042 Bucharest, Romania\\
 $^{3}$Faculty of  Physics,
University of Bucharest, \\P.O. Box MG 11, RO-077125 Bucharest,
Romania \\
and\\
Institute of Mathematics of the Romanian Academy,\\ P.O. Box
1-764, RO-014700 Bucharest, Romania}

\date{}
\maketitle

\begin{abstract}
It is proved that for general, not necessarily periodic quasi one
dimensional systems, the band position operator corresponding to
an isolated part of the energy spectrum has discrete spectrum and its
eigenfunctions have the same spatial localization  as the corresponding
spectral projection. As a consequence, an eigenbasis  of the band
position operator provides a basis of  optimally localized
(generalized)
Wannier functions for quasi one dimensional systems, and this proves the strong Marzari-Vanderbilt conjecture. If the system has
some translation  symmetries (e.g. usual translations,
 screw transformations), they are "inherited" by  the Wannier
basis.
\end{abstract}

\section{Introduction}

Wannier functions (WF) were introduced by Wannier in 1937 \cite{W} as bases
in  subspaces of states corresponding to energy bands in solids, bases
consisting of exponentially localized functions (localized orbitals).
For periodic crystals they are defined as Fourier transform of Bloch
functions of the corresponding bands. Since then WF proved to be a key tool
in quantum theory of solids as they provide a tight binding description of
the electronic band structure of solids. At the conceptual level they lay at
the foundation of all effective mass type theories
 e.g the famous Peierls-Onsager substitution describing the dynamics of Bloch
electrons in the presence of an external magnetic field (see e.g.\cite{N2}and
references therein). At the quantitative level, especially after the seminal
paper by Marzari and Vanderbilt \cite{MV}, WF become an effective tool in {\em ab initio}
computational studies of electronic properties of materials. Moreover during the last decades
WF proved to be an essential ingredient in the study of low dimensional nanostructures such as
linear chains of atoms, nanowires, nanotubes etc (see e.g. \cite{CMSN},\cite{CBRM}).
In particular WF are essential for most formulations of transport phenomena using real
space Green's function method based on Landauer-B\"uttiker formalism both at
rigorous \cite{CJM} and computational levels \cite{N},\cite{CMSN}.

A few remarks are in order here. The first one is that realistic low dimensional
systems are not strictly one (two) dimensional but rather quasi one (two) dimensional
and one has to take into account the (restricted) motion along perpendicular directions.
This adds specific features as for example the screw symmetry \
in nanotubes and nanowires absent in strictly one dimensional systems.
The second one is that realistic systems, due to the presence of defects, boundaries,
randomness etc,  do not have usually full translation symmetry  and this ask for a theory
of WF not based on Bloch formalism. Finally let us remind that contrary to a widespread
opinion (see e.g. the discussion in \cite{N2}) that WF always exist for isolated band in
solids this is not true. More precisely, in more than one dimension there are subtle
topological obstructions  and these are related to the QHE \cite{Th}, \cite{CNP}, \cite{BPCMM}: a band
for which WF are known to exist gives no contribution to the quantum Hall current. It is
then crucial to have rigorous proofs of the existence of exponentially localized WF.

For one dimensional periodic systems the existence of exponentially localized WF has been
proved by Kohn in his  classic paper \cite{K} about analytic structure of Bloch functions.
An extension of  Kohn analysis to quasi one dimensional systems has been done recently by
Prodan \cite{Pr}. As for higher dimensions
 it was known since the work by des Cloizeaux \cite{dC1} \cite{dC2} that
  there  are  obstructions  to the existence of exponentially localized WF and that these
obstructions are of topological origin (more precisely as explicitly stated in \cite{N1}
these obstructions are connected to the topology of a vector  bundle of orthogonal projections).
The fact that  for simple bands of  time reversal invariant systems the obstructions are absent
was proved by des Cloizeaux \cite{dC1} \cite{dC2} under the additional condition of the existence of centre of
inversion and by Nenciu \cite{N1} in the general case. While the proofs in \cite{dC1} \cite{dC2},
\cite{N1} did not use the vector bundle theory it was suggested  in \cite{N2},\cite{NN2}
that the characteristic classes theory in combination with some deep results in the theory of
analytic functions of several complex variables  (Oka principle) can be
 used to give alternative proof of the above results and to extend them to composite bands
of time reversal symmetric systems.
This has been substantiated recently in \cite{P}, \cite{BPCMM} where the existence of
exponentially localized Wannier functions has been proved for composite bands of  time
reversal symmetric systems in two and three dimensions settling in the affirmative  a long
standing conjecture. In conclusion the situation is
satisfactory as far as periodic time reversal symmetric Hamiltonians are considered
(as already mentioned for Hamiltonians which are not time reversal symmetric
exponentially localized Wannier functions might not exists).

As already said above both the theory and applications of Wannier
functions boosted since Marzari and Vanderbilt \cite{MV},
introduced studied and proposed methods to compute the so called
maximally localized Wannier functions (MLWFs) defined by the fact
that they minimize the position mean square deviation. It was
conjectured in \cite{MV} that they can be chosen to be real
functions and that they have "optimal" exponential localization in
the sense that they have the same exponential localization as the
integral kernel of the projection operator of the corresponding
band.
 MLWFs proved to be an
invaluable tool in the theory of electronic properties of periodic
media especially in the modern theory of electronic
polarizability (see e.g. \cite{WDRV} and references therein).

In the one dimensional case the theory of MLWFs is much more
developed. It is known \cite{MV} that MLWFs are identical to the
eigenfunctions of the "band position" operator and then they are
unique (up to uninteresting phases) and can be chosen to be real
functions. Moreover the phases of the corresponding Bloch
functions are related to the parallel transport procedure
\cite{MV}, \cite{BW}. Recently a detailed study of Wannier
functions, including their exponential decay, emphasizing the
difference between the cases with and without inversion symmetry
appeared in \cite{BN}. In the same paper there are pointed out situations in which the 
Wannier functions could decay slower than the kernel of the projector, which shows that choosing the optimal phase is not a trivial task. Our results show
that that by choosing the right phase one must always obtain an optimal decay.

Motivated by  the great interest in nonperiodic structures much  effort has been devoted
to extend the results about existence of exponentially localized bases for isolated bands
in nonperiodic systems. The basic difficulty stems from the fact that for nonperiodic
systems one cannot define Wannier functions as Fourier transforms of the Bloch functions.
One way out of the difficulty is to start from the periodic case or tight-binding limit
where the Wannier functions are known to exist and and use perturbation or ``continuity''
arguments. The basic idea is that since the obstructions are of topological origin the
existence of exponentially localized WF is stable against perturbations. Indeed along
these lines  it has been possible to prove the existence of (generalized) WF for a
variety of nonperiodic systems \cite{KO}, \cite{RK}, \cite{GK}, \cite{N2}, \cite{NN2}.
Since in the periodic case the obstructions to the existence of exponentially localized WF
are absent \cite{dC1},\cite{dC2},\cite{N1} in one dimension it was   naturally to
conjecture \cite{NN2},\cite{Ni} that in one dimension  WF exist for all isolated bands
irrespective of  periodicity properties.

The first problem to be solved was to find an alternative definition of WF. The basic
idea goes back to Kivelson \cite{Ki},  who proposed to {\em define} the generalized WF
as eigenfunctions of the ``band position'' operator.
 To substantiate the idea one has to prove that the band position operator is self-adjoint,
has discrete spectrum and its eigenfunctions are exponentially localized.
 For the particular case of a periodic one dimensional crystal with one defect Kivelson proved
that the eigenfunctions of the band position operator are indeed exponentially localized
and asked for a general proof. In the general case, by a bootstrap argument, Niu  \cite{Ni} argued
that the
eigenfunctions of the band position operator (if they exist) are at least polynomially localized.
In full generality the fact that for all isolated parts of the spectrum the band position
operator is self-adjoint, has discrete spectrum and its eigenfunctions are exponentially
localized has been proved in \cite{NN3}.

In this paper we extend  the results  in \cite{NN3} to quasi
one-dimensional systems i.e. three dimensional systems for which
the motion extends to infinity only in one direction. In addition
we add the result (which is new even in the strictly one
dimensional case) that (see Theorem \ref{main} below for details)
the ``density"  of WF is uniformly bounded. While the main ideas
of the proof are the same as in \cite{NN3} there are major
differences both at the technical  and physical level. In
particular for quasi one dimensional systems with screw symmetry
the constructed WF inherits this symmetry a property which is very
useful in computational applications. Finally let us point out
that as in the periodic case, generalized WF defined as
eigenfunctions of the band position operator have very nice
properties e.g. they are (up to uninteresting phases) uniquely
defined and for real (i.e. time reversal invariant) Hamiltonians
they can be chosen to be real functions and this solves for the
general quasi one dimensional case the ``strong conjecture'' in
 Section V. of \cite{MV}. As for their exponential localization we
 have the following "optimality" result (see Proposition \ref{optimal} 
for a precise statement)
 which seems to be new even
 in the one dimensional periodic case: the
 eigenfunctions of the band position operator have 
the same exponential localization as the integral kernel of the projection
 operator of the corresponding band.

\section{The results}

Consider in $L^{2}(\bR^{3})$ the following Hamiltonian describing a
particle subjected to a  scalar potential $V$:
\begin{equation}\label{ham}
H=\bP^{2}+V,
\; \; \; {\bf P}=-i\nabla,\; \;\;
\sup_{{\bf x} \in {\bR }^3} \int_{\vert {\bf x-y} \vert\leq 1}
\vert V({\bf y}) \vert^2 d{\bf y} < \infty
\end{equation}
which, as is well known (see \cite {RS}), is essentially self-adjoint
on $C_0^{\infty}({\bR}^3)$.  We have already said in the introduction
that we are interested in potentials $V$ which tend to zero as the
distance from the $Ox_{1}$ axis tends to infinity. Let us now be more
precise. The notation $\bx = (x_{1},\bx_{\perp})$ will be used
throughout the paper. For any $R>0$, define:
\begin{equation}\label{l2locu}
I_{V}(R):=\sup_{x_{1}\in \bR ,|\bx_{\perp}| \geq R}  \int_{\vert {\bf x-y} \vert\leq 1}
\vert V({\by}) \vert^2 d{\by}.
\end{equation}
The decay assumption for $V$ will be:
\begin{equation}\label{confv}
\lim_{R\rightarrow \infty}I_{V}(R)=0.
\end{equation}
It is easy to see that $[0,\infty)\subset \sigma(H)$ (using a Weyl
sequence argument), thus the only region where $H$ might have an
isolated spectral island is below zero.
Now suppose that $\sigma_0$ is such an isolated part of the spectrum
and define:
\begin{equation}\label{E+}
-E_{+}:= \sup \{ E:\; E \in \sigma_{0}\} < 0.
\end{equation}
If $\Gamma$ is a positively oriented contour of finite
 length enclosing $\sigma_0$, then the spectral subspace corresponding
 to $\sigma_0$ is:
\begin{equation}\label{subs}
{\cal K}\;:={\rm Ran}(P_0),\quad P_0= \frac{i}{2\pi}
\int_{\Gamma}(H-z)^{-1}dz.
\end{equation}

At a heuristic level, due to the fact that the wave packets from
$\cK$ cannot propagate in the classically forbidden region (see
\eqref{E+} and \eqref{confv}), at negative energies the motion is
confined near the $Ox_{1}$ axis, i.e. the system has a quasi one
dimensional behavior.

\subsection{The technical results}

The following proposition states the "localization"
 properties of $P_0$. On one hand, this give a precise meaning to the
 previously discussed quasi one dimensional character,
and on the other hand it provides some key ingredients to the proof of
 exponential localization of eigenfunctions of the band position
 operator.

Let $a\in \bR$, and let $\langle X_{\parallel, a}\rangle$ be the
multiplication operator corresponding to:
\begin{equation}\label{ga}
g_a(\bx):=\sqrt{(x_1-a)^2 +1},
\end{equation}
and $\langle X_{\perp}\rangle$ be multiplication operator given by:
\begin{equation}\label{gp}
g_{\perp}(\bx):=\sqrt{|\px|^2 +1}.
\end{equation}
\begin{proposition}\label{locP0}
There exist $\alpha_{\parallel}>0$,  $\alpha_{\perp}>0$, $M<\infty$ such
that:
\begin{equation}\label{expP0}
\sup_{a\in \bR} \parallel e^{\paralpha \parX}P_0 e^{-\paralpha \parX}
\parallel \leq M,\quad {\rm and}
\end{equation}
\begin{equation}\label{confP0}
 \parallel e^{\peralpha \perX}P_0 e^{\peralpha \perX}
\parallel \leq M.
\end{equation}
\end{proposition}
\noindent The proof of Proposition \ref{locP0} will also give values for $\paralpha$
and $\peralpha$. In particular $\peralpha$ can be any number
strictly smaller than $\sqrt{E_+}$. 

\vspace{0.5cm}

We now can formulate the main technical result of this paper. To emphasize its
generality we stress that its proof only uses the decay condition
(\ref{confv}) and the existence of an isolated part of the
spectrum satisfying (\ref{E+}).
\begin{theorem}\label{main}
Let $X_{\parallel}$ be the operator of multiplication with $x_1$ in
$L^2(\bR^3)$ and
consider in $\cK$ the operator
\begin{equation}\label{hX}
\hat{X}_{\parallel}:=P_{0}X_{\parallel}P_{0}
\end{equation}
defined on
$
{\cal D}(\hX)={\cal D}(X_{\parallel})\cap{\cal K}.
$
Then

\noindent {\rm i}. $\hX$ is self-adjoint on ${\cal D}(\hat{X})$;

\noindent {\rm ii}. $\hX$ has purely discrete spectrum;

\noindent {\rm iii}. Let $g \in G := \sigma(\hX)$ be an eigenvalue,
 $m_g$ its multiplicity, and $\{W_{g,j}\}_{1\leq j\leq
   m_g}$ an orthonormal basis in the eigenspace of
 $\hat{X}$ corresponding to $g$. Then for all $\beta \in [0,1]$,
there exists $ M_1<\infty$ independent of $ g$, $j $ and
$\beta $ such that:
\begin{equation}\label{locW}
\int_{\bR^3}e^{2(1-\beta) \paralpha \vert x_1-g\vert}
e^{2\beta \peralpha \vert \px\vert}\vert
 W_{g,j}(\bx)\vert ^2
d\bx \leq M_1,
\end{equation}
where $\paralpha$ and $\peralpha$ are the same exponents as those provided
by the proof of Proposition \ref{locP0};

\noindent {\rm iv}. Let $a \in \bR$ and $L \geq 1$. 
Denote by $N(a,L)$ the  total multiplicity of the spectrum of $\hX$ 
contained in $[a-L,a+L]$.
Then there exists $M_2 < \infty$ such that
\begin{equation}\label{density}
N(a,L) \leq M_2 \cdot L.
\end{equation}
\end{theorem}

Finally, we turn to the question of optimal localization properties of
our Wannier functions. Theorem \ref{main} provides an optimal
exponential decay on the transverse direction, but in the
parallel direction it only implies a decay which is bound by the
maximal decay of the
resolvent in the gap. The conjecture on optimal exponential decay, as
stated in Section V of \cite{MV}, is whether
 the $ W_{g,j}$'s have the same exponential decay as the  integral kernel 
$\cP_0(\bx,\by)$ of $P_0$ (which can be larger than the
maximal decay of the resolvent in the gap; we are
  indebted to one of the referees for pointing this to us). 
Concerning this issue, we have the following result showing the
optimality of the "parallel" decay of $
  W_{g,j}$ at the exponential level.

\begin{proposition}\label{optimal}
Assume  that for all $\alpha<\alpha_0$ we are given an {\rm a priori} bound
 \begin{equation}\label{decP}
 \sup_{a\in \bR} \parallel e^{\alpha \parX}P_0 e^{-\alpha \parX}
\parallel <\infty.
\end{equation}
Then
for all $\alpha<\alpha_0$ there exists $M_1(\alpha)$, independent
of $g$ and $j$, such that
\begin{equation}\label{optW}
\int_{\bR^3}e^{2 \alpha \vert x_1-g\vert} \vert
 W_{g,j}(\bx)\vert ^2
d\bx \leq M_1(\alpha).
\end{equation}
\end{proposition}

\vspace{0.5cm}

\noindent{\bf Remark}. Here $\alpha_0$ is the "exact" exponential
decay of $\cP_0(\bx,\by)$. In certain particular periodic cases 
one might obtain a power-like asymptotic behavior of 
$e^{\alpha_0|x_1-y_1|}\cP_0(\bx,\by)$ in the variables $x_1,y_1$. 
We cannot say anything about an eventual asymptotic behavior of 
$e^{\alpha_0 \vert x_1-g\vert}W_{g,j}(\bx)$. But due to the generality
of the setting, we consider our result to be optimal.

\subsection{Further properties of the Wannier basis}

We come now to the case when $V$ (hence $H$) has additional
symmetries. The point here is that although the Wannier functions are
not eigenfunctions of $H$, one would like them to inherit in some sense
the symmetries of $H$. The reason is that usually the Wannier basis is
used in order to write down an effective Hamiltonian in $\cK$, and one would
like this effective Hamiltonian to inherit as much as possible the
symmetries of $H$.

First we comment on time reversal invariance.
Since $V(\bx) $ is real, $H$ commutes with the anti-unitary operator
induced by complex conjugation. It follows (see (\ref{subs})) that
$P_0$ and $\hX$ are also real, thus the eigenfunctions of $\hX$ can be
chosen to be real. Hence Theorem \ref{main} provides us with a Wannier basis
which is time reversal invariant.

Second we consider the so called "screw-symmetry" along the $Ox_1$-axis,
of much interest in the physics of carbon nanotubes. Namely, writing
\begin{equation}\label{cil}
\px =(r,\theta),\quad r\geq 0,\; \theta \in [0,2\pi),
\end{equation}
one assumes that for some  $\theta_0 \in [0,2\pi)$ we have:
\begin{equation}\label{screwv}
V(x_1,r,\theta )=V(x_1+1,r,\theta+\theta_0 ).
\end{equation}
Here $\theta+\theta_0$ has to be understood modulo $2\pi$.
Defining the  screw-symmetry operators $T_n^{\theta_0}$ by:
\begin{equation}\label{Tn}
(\Tn f)(x_1,r,\theta ):=f(x_1-n,r,\theta-n\theta_0 ),
\end{equation}
one has a  (unitary!) representation of $\mathbb{Z}$ in $L^2(\bR^3)$.
Taking into account (\ref{screwv}) and the fact that $[-\Delta,\Tn]=0$
(use cylindrical coordinates to prove this), one
obtains:
\begin{equation}\label{HTn}
[H,\Tn]=0,
\end{equation}
and then from functional calculus and (\ref{subs}):
\begin{equation}\label{P0Tn}
[P_0,\Tn]=0.
\end{equation}
In particular, this implies that the family $\{\Tn\}_{n\in
  \mathbb{Z}}$ induces a unitary representation of $\mathbb{Z}$ in
$\cK$. Moreover, from (\ref{hX}) and
\ref{P0Tn}) one obtains:
\begin{equation}\label{TnX}
[\Tn,\hX]=n\Tn.
\end{equation}
Let $p <\infty$ be the number of eigenvalues of $\hX$ in the interval
$[0,1)$, and let $\{g_j\}_{j=1}^p$ be the distinct eigenvalues (each
with multiplicity $m_j <\infty$). We have:
\begin{equation}\label{Wuc}
\hX W_{g_j,\alpha_j}=g_j W_{g_j,\alpha_j}, \;\alpha_j= 1,2,...,m_{g_j}.
\end{equation}
From (\ref{TnX} ) and (\ref{Wuc}) one obtains that for all
$g_j,\;\alpha_j,\;n \in \mathbb{Z}$:
\begin{equation}\label{Wn}
\hX \Tn  W_{g_j,\alpha_j}= (g_j+n)\Tn  W_{g_j,\alpha_j}.
\end{equation}
Conversely, for every other $g \in \sigma (\hX )$, choose an
eigenvector $W_{g}$. We can find $n\in \mathbb{Z}$ such that  $g+n\in
[0,1)$. Since $\hX \Tn  W_{g}= (g+n)\Tn  W_{g}$, it means that $g+n$
must be one of the $g_j$'s considered above. Therefore we proved the
following corollary:
\begin{corollary}\label{screwW}
The spectrum of $\hX$ consists of a union of $p$ ladders:
\begin{equation}\label{spectX}
G=\cup_{j=1}^p G_j, \quad G_j=\{g:\; g=g_j+n,\;n\in \mathbb{Z} \},\quad j\in\{1,2,...,p\},
\end{equation}
and an orthonormal basis in $\cK$ can be chosen as:
\begin{align}\label{Wbasis}
 &W_{n,g_j,\alpha_j}:= W_{g_j+n,\alpha_j}:=\Tn W_{g_j,\alpha_j}, \\
& n \in \mathbb{Z},\; j\in \{1,2,...,p\},\;\alpha_j\in\{
1,2,...,m_{g_j}\}.\nonumber
\end{align}
\end{corollary}

\vspace{0.5cm}

It is interesting to express the effective Hamiltonian
$P_0HP_0$ as an infinite matrix with the help of the Wannier basis.
For notational simplicity we relabel the pair $(g_j,\;\alpha_j)$ as
$l\in \{1,2,...,N_c=\sum_{j=1}^p m_{g_j}\}$ and write the Wannier basis as
 $\{W_{n,l}\}_{n \in \mathbb{Z},\;l\in \{1,2,...,N_c\}}$. Note that
 $N_c$ is nothing that the number of Wannier functions per unit cell $[0,1)$. Let
\begin{equation}
h_{l,k}^{\theta_0}(m,n):=\langle W_{m,l},HW_{n,k}\rangle.
\end{equation}
The important fact is that in spite of a
rotation with an angle $\theta_0$ for which it might happen that
$\frac{\theta_0}{2\pi}$ to be irrational, from \eqref{HTn} and
(\ref{Wbasis}) one obtains (with the usual abuse of notation):
\begin{equation}\label{transl}
h_{l,k}^{\theta_0}(m,n)=h_{l,k}^{\theta_0}(m-n).
\end{equation}
Then a standard computation gives the effective Hamiltonian as an
operator in $(l^2)^{N_c}$ which is of {\em standard translation invariant}
tight binding type:
\begin{equation}\label{heff}
(h_{eff}^{\theta_0}\phi)_l(m):=\sum_{k,n}h_{l,k}^{\theta_0}(m-n)\phi_k(n).
\end{equation}
This is another consequence of the quasi one-dimensional character of the
motion for negative energies. More precisely, it  reflects the fact that for arbitrary
values of $\theta_0$, since $\Tn$ is a unitary
representation of $\mathbb{Z}$, one can still develop a Bloch type analysis but with
a more complicated form of "Bloch" functions:
\begin{equation}\label{bloch}
\Psi_k(\bx) = e^{ikx_1}u_k (\bx),\; u_k (\bx)=\Tn u_k (\bx).
\end{equation}
However, due to the complicated symmetry of the resulting Bloch functions (which does
not allow to represent the fiber Hamiltonian as a differential
operator on the unit cell with "simple" boundary conditions), the
analysis gets much harder. The Bloch analysis reduces to the standard
one (with a larger unit cell) for rational
values of  $\frac{\theta_0}{2\pi}$.

\section{Proofs}

This section is devoted to the proof of Proposition \ref{locP0}, 
Theorem \ref{main} and Proposition \ref{optimal}. A certain number
of unimportant finite positive constants appearing during the
proof will be denoted by $M$.

One of the key ingredients in the proofs is the exponential decay
of the integral kernel of the resolvent of Schr\"odinger
operators. This is an elementary result in the Combes-Thomas-Agmon
theory of weighted estimates. We summarize the needed result in:
\begin{lemma}\label{CTAL}
Let $W$ be a potential such that
$\sup_{{\bf x} \in {\bf R }^3} \int_{\vert {\bf x-y} \vert\leq 1}
\vert W({\bf y}) \vert^2 d{\bf y} < \infty$. Define $K:=\bP ^{2}+
W(\bx)$ as an operator sum, and let $h$ be a real function
satisfying:
\begin{align}\label{prima1}
h\in C^{\infty}(\bR^3),\quad \sup_{\bx\in\bR^3}\{
|\nabla h(\bx)|+ |\Delta  h(\bx)|\}= m <
\infty.
\end{align}
Fix $z\in \rho(H)$. Then there exists $\alpha_{z }>0$ such that
\begin{equation}\label{prima11}
\Vert e^{\alpha_{z}h}
(K-z)^{-1}e^{-\alpha_{z}h}\Vert \leq M,
\end{equation}
\begin{equation}\label{adoua11}
\Vert e^{\alpha_{z}h}
P_j(K-z)^{-1}e^{-\alpha_{z}h}\Vert \leq M,
\end{equation}
where $P_j=-i\frac{\partial}{\partial x_j }$, $j\in\{1,2,3\}$.
\end{lemma}
Without giving the details of the proof of Lemma \ref{CTAL}, for later
use we write down a key identity in \eqref{CTA}: under the condition
\begin{equation}\label{CTAcond}
1+\alpha_{z} (\pm i\bP\cdot \nabla h \pm i \nabla h\cdot \bP
-\alpha_{z}|\nabla h|^2)
(K-z)^{-1} \quad {\rm invertible}
\end{equation}
one has
\begin{align}\label{CTA}
 & e^{\pm \alpha_{z}h}
(K-z)^{-1}e^{\mp \alpha_{z}h}\\
&= (K-z)^{-1}[1+\alpha_{z} (\pm i\bP\cdot
\nabla h \pm i \nabla h\cdot \bP
-\alpha_{z}|\nabla h|^2)
(K-z)^{-1}]^{-1}. \nonumber
\end{align}
Then (\ref{CTAcond}) holds true if for example $\alpha_{z}>0$ is small
enough.

\vspace{0.5cm}

\subsection{Proof of Proposition \ref{locP0}}
Take $\Gamma$ in
(\ref{subs}) a
contour of finite length enclosing $\sigma_0$ and satisfying
\begin{equation}\label{Gamma}
\dist (\Gamma, \sigma (H))=\frac{1}{2}\dist (\sigma_0 , \sigma
(H)\setminus \sigma_0 ).
\end{equation}
Then since $|\nabla g_a | \leq 1$, $|\Delta g_a |^2 \leq 2$, the
estimate (\ref{expP0}) follows directly from Lemma \ref{CTAL}  by taking
$\paralpha$ sufficiently small such that for all $z \in\Gamma $:
$$
\Vert \paralpha (i\bP\cdot \nabla g_a +i \nabla g_a \cdot \bP
-\paralpha |\nabla g_a|^2)
(K-z)^{-1} \Vert \leq b <1.
$$
We now prove (\ref{confP0}). If $R>0$, define:
\begin{equation}\label{HR}
H_R= -\Delta +(1-\chi_R)V,
\end{equation}
where
\begin{equation}\label{carR}
\chi_R(\bx )=\left\{\begin{array}{rlc}
1&\mbox{for}&\vert \px \vert \leq R\\
0&\mbox{for}&\vert \px \vert > R\end{array}\right. .
\end{equation}
From (\ref{confv}) it follows that
\begin{displaymath}
\lim_{R\rightarrow \infty} \inf \sigma (H_R) =0.
\end{displaymath}
In particular, for sufficiently large $R$,  $(H_R-z)^{-1}$ is analytic
inside $\Gamma$. Since $H-H_R=\chi_R V$, then using resolvent
identities we obtain:
\begin{align}\label{HHR}
&(H-z)^{-1}=(H_R-z)^{-1}\\
&-(H_R-z)^{-1}\chi_R V(H_R-z)^{-1}+
(H_R-z)^{-1}\chi_R V(H-z)^{-1}\chi_R V(H_R-z)^{-1}.\nonumber
\end{align}
From (\ref{subs}), (\ref{HHR}) and the fact that  $(H_R-z)^{-1}$ is analytic
inside $\Gamma$ one has
\begin{equation}\label{P0R}
P_0= \frac{i}{2\pi}
\int_{\Gamma}(H_R-z)^{-1}\chi_R V(H-z)^{-1}\chi_R V(H_R-z)^{-1}.
\end{equation}
Notice that for all $\alpha >0$:
\begin{equation}\label{RV}
\sup_{{\bf x} \in {\bf R }^3} \int_{\vert {\bf x-y} \vert\leq 1}
\vert (e^{\alpha \pg}\chi_R V)({\bf y}) \vert^2 d{\bf y} < \infty.
\end{equation}
Take now $\peralpha >0$ such that \ref{CTAcond} holds true for all $z
\in \Gamma$, $K=H_R$, $h=\pg$ and $\alpha_z=\peralpha $. That is let
us suppose that
\begin{equation}\label{acincea6}
1+\peralpha (\pm i\bP\cdot \nabla \pg \pm i \nabla \pg\cdot \bP
-\peralpha |\nabla \pg|^2)
(H_R-z)^{-1} \quad {\rm is}\:\: {\rm invertible}
\end{equation}
uniformly on $\Gamma$.  Then we can rewrite
$P_0$ as:
\begin{align}\label{prima6}
P_0&=e^{-\peralpha \perX}\left \{\frac{i}{2\pi}\int_{\Gamma}
\left [e^{\peralpha \perX}(H_R-z)^{-1}e^{-\peralpha \perX}\right
]\right . \nonumber \\
&\left [e^{\peralpha \pg}\chi_R V(H-z)^{-1}\right ]\;
\left [e^{\peralpha \pg}\chi_R V
(H_R-z)^{-1}\right ] \\
&\left . \left [1+\peralpha (-i\bP\cdot \nabla \pg -i \nabla \pg\cdot \bP
-\peralpha |\nabla \pg|^2)
(H_R-z)^{-1}\right ]^{-1}dz\right \} e^{-\peralpha \perX}.\nonumber
\end{align}
 Due to \eqref{RV} the operator under the integral sign is uniformly bounded in
 $z$ and the proof of Proposition \ref{locP0} is finished provided we
 can show why we can choose $\peralpha$
 as close to $\sqrt{E_+}$ as we want. The argument is as follows. Choose
 $0\leq \peralpha <\sqrt{E_+}$. Choose a  contour $\Gamma$ which is very close
 to $\sigma_0$, at a distance $\delta>0$, infinitesimally small.
Using the spectral theorem (or in this
 case the Plancherel theorem), there exists $\delta$ small enough such that the
 following estimates hold true:
\begin{align}\label{adoua6}
\sup_{z\in\Gamma}\left \Vert (\bP^2-z)^{-1}\right \Vert \leq {\rm const},\quad \sup_{z\in\Gamma}\max_{j\in\{1,2,3\}}\left \Vert P_j(\bP^2-z)^{-1}\right \Vert \leq
{\rm const}.
\end{align}
Hence we can find $\delta$ small enough and $R$
large enough such that the operator in \eqref{acincea6} is invertible
if
\begin{equation}\label{asasea6}
1+\peralpha (\pm i\bP\cdot \nabla \pg \pm i \nabla \pg\cdot \bP
-\peralpha |\nabla \pg|^2)
(\bP^2-\Re(z))^{-1}\quad {\rm is}\:\: {\rm invertible}
\end{equation}
uniformly on $\Gamma$.
Now the operator in \eqref{asasea6} is invertible if
\begin{align}\label{atreia6}
1&\pm i\peralpha (\bP^2-\Re(z))^{-\frac{1}{2}}
(\bP\cdot \nabla h +\nabla h\cdot
\bP)(\bP^2-\Re(z))^{-\frac{1}{2}} \nonumber \\
&-\peralpha^2(\bP^2-\Re(z))^{-\frac{1}{2}}|\nabla h|^2
(\bP^2-\Re(z))^{-\frac{1}{2}}
\end{align}
is invertible (by a resummation of the Neumann series and analytic
continuation). Now assume that uniformly on $\Gamma$ we have:
$$0<\peralpha^2(\bP^2-\Re(z))^{-\frac{1}{2}}|\nabla h|^2
(\bP^2-\Re(z))^{-\frac{1}{2}}\leq \frac{\peralpha^2}{-\Re(z)}<1,$$
which can be achieved if $\peralpha^2<E_+$ and $\delta$ is chosen to be
small enough. Define
$$S:=\left (1- \peralpha^2(\bP^2-\Re(z))^{-\frac{1}{2}}|\nabla h|^2
(\bP^2-\Re(z))^{-\frac{1}{2}}\right)^{-\frac{1}{2}},$$
and
$$T=T^*:=S(\bP^2-\Re(z))^{-\frac{1}{2}}
(\bP\cdot \nabla h +\nabla h\cdot
\bP)(\bP^2-\Re(z))^{-\frac{1}{2}}S.$$
Then the operator in \eqref{atreia6} is invertible if $1\pm i\peralpha
T$ is invertible, which is always the case:
$$(1\pm i\peralpha T)^{-1}=(1\mp i\peralpha
T)(1+\peralpha^2T^2)^{-1}.$$
Therefore Proposition \ref{locP0} is proved. \qed

\subsection{ Proof of Theorem \ref{main}}
\noindent {\it Proof of} (i). First we recall an older result
(see e.g. \cite{A, N2, NN1}), according to which the commutator
$[X_{\parallel},P_0]$
defined on ${\cal D}(X_{\parallel})$
has a bounded closure on $L^2 (\bR^3)$. We seek an approximate
resolvent of $\hX$ by defining  for $\mu > 0$ the operator
\begin{equation}\label{ares}
\hat{R}_{\pm \mu}=P_0 (X_{\parallel} \pm i\mu)^{-1}P_0.
\end{equation}
Since one can rewrite $\hat{R}_{\pm \mu}$ as
$$
\hat{R}_{\pm \mu}=(X_{\parallel} \pm i\mu)^{-1}P_0+
(X_{\parallel} \pm i\mu)^{-1}[X_{\parallel},P_0]
(X_{\parallel} \pm i\mu)^{-1}P_0
$$
it follows that
$\hat{R}_{\pm \mu}{\cal K}\subset \it D (\hX)$
and by a straightforward computation (as operators in ${\cal K}$)
\begin{equation}
(\hX \pm i\mu)\hat{R}_{\pm \mu}=
P_0 (X_{\parallel} \pm i\mu)P_0(X_{\parallel} \pm i\mu)^{-1}P_0=
1_{\cal K} + \hat{A}_{\pm \mu}
\end{equation}
with
\begin{equation}
\hat{A}_{\pm \mu}=P_0[X_{\parallel},P_0 ] (X_{\parallel} \pm i\mu)^{-1}P_0.
\end{equation}
 Since  $[X_{\parallel},P_0]$ is bounded and $\Vert (X_{\parallel} \pm
 i\mu)^{-1}\Vert \leq \frac{1}{\mu}$, it follows that for
sufficiently large $\mu$:
\begin{equation}
\Vert\hat{A}_{\pm \mu}\Vert \leq \frac{1}{2}.
\end{equation}
Then again as operators in ${\cal K}$:
\begin{equation}\label{surj}
(\hat X \pm i\mu)\hat{R}_{\pm \mu}
(1_{\cal K} + \hat{A}_{\pm \mu})^{-1}=1_{\cal K}
\end{equation}
This implies that $\;\;\hat X \pm i\mu\;\;$is surjective
on $\;\;\hat{R}_{\pm \mu}
(1_{\cal K} + \hat{A}_{\pm \mu})^{-1}{\cal K}\subset {\it D}(\hat X)\;$.
By the fundamental criterion of self-adjointness \cite{RS}
$\hat X$ is self-adjoint in ${\cal K}$ on ${\cal D}(\hat X)$.
In addition, from \eqref{surj} one obtains the following formula for the
resolvent of $\hX$:
\begin{equation}\label{hXres}
(\hX \pm i\mu)^{-1}=\hat{R}_{\pm \mu}
(1_{\cal K} + \hat{A}_{\pm \mu})^{-1}.
\end{equation}

\vspace{0.5cm}

\noindent {\it Proof of} (ii).
We will show that $\hat{R}_{\pm \mu}$ is compact in ${\cal K}$ which
implies (see \eqref{hXres}) that $\hX$ has compact resolvent, thus
purely discrete spectrum. Consider a cut-off function $\phi_N$ which
equals $1$ if $|\bx|\leq N$ and is zero if $|\bx|\geq 2N$.
For  $N\geq 1 $ we can decompose:
\begin{equation}\label{prima10}
\hat{R}_{\pm \mu}=P_0 (X_{\parallel} \pm i\mu)^{-1}\phi_NP_0+
P_0 (X_{\parallel} \pm i\mu)^{-1}(1-\phi_N)P_0.
\end{equation}
Writing $$\phi_NP_0=\{\phi_N(\bP^2+1)^{-1}\}\{(\bP^2+1)P_0\}$$ we see
that $\phi_NP_0$ is compact (even Hilbert-Schmidt)
in $L^2(\bR^3)$ (the first factor is Hilbert-Schmidt while the
second one is bounded). Now if $0<\alpha$ is small enough, we know
that $e^{\alpha g_\perp}P_0$ is bounded (see \eqref{confP0}). Since
$$\lim_{N\to \infty}\left \Vert (X_{\parallel} \pm i\mu)^{-1}(1-\phi_N)
e^{-\alpha g_\perp}\right \Vert =0,$$
we have shown:
$$\lim_{N\to \infty}\left \Vert \hat{R}_{\pm \mu}-
P_0 (X_{\parallel} \pm i\mu)^{-1}\phi_NP_0\right \Vert =0,$$
thus $\hat{R}_{\pm \mu}$ equals the norm limit of a sequence of compact
operators, therefore it is compact. Accordingly, since the
self-adjoint operator $\hX$ has compact resolvent
it has purely discrete spectrum \cite{RS}:
\begin{equation}
\sigma(\hX)=\sigma_{disc}(\hX)=: G,
\end{equation}
and the proof of the second part of Theorem \ref{main} is finished.

\vspace{0.5cm}

\noindent{\it Proof of} (iii). Now we will consider the exponential
localization of eigenfunctions of $\hX$. Let $g \in G$ be an eigenvalue,
 $m_g$ its multiplicity and $W_{g,j},\;\; 1\leq j\leq m_g$ be
 an orthonormal basis in the eigenspace of $\hX$ corresponding to
 $g$. We shall prove that uniformly in $g$
 and $j$
\begin{align}\label{parW}
\Vert  e^{\paralpha \pargX}W_{g,j} \Vert & \leq M \quad {\rm and}
\\ \label{perW}
\Vert  e^{\peralpha \perX}W_{g,j} \Vert & \leq M.
\end{align}
 Taking \eqref{parW} and \eqref{perW} as given, one can easily
 obtain \eqref{locW} by a simple convexity argument: the function
$ f(x)= a^{1-x}b^{x}$ ; $a,b>0$ is convex on $\bR$, and for $0\leq \beta
\leq 1$ one has:
\begin{equation}
\beta e^{2 \paralpha g_a(\bx)} +(1-\beta )e^{2 \peralpha \pg }
\geq  e^{2(1- \beta)\paralpha g_a(\bx)}e^{2 \beta \peralpha \pg },
\end{equation}
which together with  \eqref{parW} and \eqref{perW} it proves
\eqref{locW} with $M_1=M^2$. Since \eqref{perW} follows directly from
\eqref{confP0} and $W_{g,j}=P_0 W_{g,j}$ we are left with
the proof of \eqref{parW}.

Although the proof of \eqref{parW} mimics closely the proof in the one
dimensional case \cite{NN3}, we give it here for completeness. In order
to emphasize the main idea of the proof let us remind one of the
simplest proofs of the exponential decay of eigenfunctions of Schr\"odinger
operators corresponding to discrete eigenvalues (assuming that the
potential $V$ is bounded and has compact support). Namely assume that for
some $E>0$ we have $(-\Delta +V+E)\Psi =0$,
which  can be rewritten as
\begin{equation}\label{Svp}
\Psi =- (-\Delta +E)^{-1}V\Psi.
\end{equation}
Since for $|\alpha|<\sqrt{E}$, $e^{\alpha |\cdot|} (-\Delta
+E)^{-1}e^{-\alpha |\cdot|}$ and $e^{\alpha |\cdot|}V$ are bounded:
$$
\Psi =- e^{-\alpha |\cdot|}\left \{e^{\alpha |\cdot|} (-\Delta
+E)^{-1}e^{-\alpha |\cdot|}\right \}(e^{\alpha |\cdot|}V)\Psi
$$
which proves the exponential localization of $\Psi$. The main idea in
proving
\eqref{parW} is to rewrite the eigenvalue equation for $\hX$ in a form
similar to \eqref{Svp} and and then to use \eqref{expP0}.

Let us start with some notation. If $b>0$ (sufficiently large) and
$a\in {\bR}$, define:
\begin{equation}\label{adoua10}
f_{a,b}(\bx):=b\;f\left (\frac{x_1-a}{b}\right )
\end{equation}
where $f$ is a real $C_0^{\infty}(\bR)$ cut-off function satisfying
$0\leq f(y)\leq 1$ and
\begin{displaymath}
f(y)=\left\{\begin{array}{rlc}
1&\mbox{for}&\vert y \vert \leq \frac{1}{2}\\
0&\mbox{for}&\vert y \vert \geq 1\end{array}\right. .
\end{displaymath}
Define the function $h_{a,b}$ by:
\begin{equation}\label{hab}
h_{a,b}(\bx):=x_1-a+if_{a,b}(\bx).
\end{equation}
Note that by construction, $ h_{a,b}$ only depends on $x_1$, and obeys:
\begin{equation}\label{invh}
\vert h_{a,b}(\bx) \vert \geq \frac{b}{2}.
\end{equation}
Moreover, its first two derivatives are uniformly bounded:
\begin{equation}\label{derh}
\sup_{\bx \in\bR^3}\sup_{a\in {\bR}}\sup_{b\geq 1} \lbrace \vert
\nabla h_{a,b}(\bx)\vert + \vert \Delta h_{a,b}(\bx)
 \vert \rbrace =  K< \infty.
\end{equation}
The eigenvalue equation for $W_{g,j}$ reads as
$P_0 (\hX-g)P_0 W_{g,j}  = 0$. Using \eqref{hab} it can be rewritten as:
\begin{equation}\label{meee}
P_0 h_{g,b}P_0 W_{g,j}=iP_0 f_{g,b}P_0 W_{g,j}.
\end{equation}
We now prove that $P_0 h_{g,b}P_0$ is invertible.
Like in the proof self-adjointness of $\hX$ we compute
\begin{equation}\label{par}
P_0 h_{g,b}^{-1}P_0 P_0 h_{g,b}P_0 =
1_{\cal K}+P_0 h_{g,b}^{-1}\left[P_0, h_{g,b}\right]P_0.
\end{equation}
The key remark is that $\left[P_0, h_{g,b}\right]$ is bounded. Indeed
we have the identity:
\begin{align}\label{P0h}
\left[P_{0},h_{g,b}\right]&=
-\frac{1}{2\pi }\int_{\Gamma}(H-z)^{-1}
\left\{ \bP \cdot \nabla h_{g,b} +
\nabla h_{g,b}\cdot \bP \right\}(H-z)^{-1}dz\nonumber \\
&=
-\frac{1}{2\pi }\int_{\Gamma}(H-z)^{-1}
\left\{ -i\Delta h_{g,b} +
2\nabla h_{g,b}\cdot \bP \right\}(H-z)^{-1}dz.
\end{align}
It follows that
  $\left[P_0, h_{g,b}\right]$ is uniformly bounded in $g\in{\bR}$ and
  $b\geq 1$ (see \eqref{derh}).
Taking into account \eqref{invh} one obtains that the operator
\begin{equation}\label{B}
\hat{B}_{g,b}=P_0 h_{g,b}^{-1}\left[P_0, h_{g,b}\right]P_0
\;\;:\;\;{\cal K}\rightarrow{\cal K}
\end{equation}
satisfies
\begin{equation}\label{B12}
\Vert\hat{B}_{g,b}\Vert \leq \frac{1}{2}
\end{equation}
if $b\geq b_0$ for some large enough $b_0 <\infty$.
It follows that $1+\hat{B}_{g,b}$ is invertible and then the
eigenvalue equation (see \eqref{meee}, \eqref{par} and \eqref{B})
 takes the form
\begin{equation}\label{eef}
W_{g,j}=i \left(1+\hat{B}_{g,b}\right)^{-1}
P_0 h_{g,b}^{-1}P_0 f_{g,b}P_0 W_{g,j}
\end{equation}
which is the analog of \eqref{Svp}. By construction  (see the
definition of $f_{g,b}$ in \eqref{adoua10}):
\begin{displaymath}
\Vert e^{\paralpha \pargX}f_{g,b}\Vert \leq
be^{\paralpha (b+1)}.
\end{displaymath}
Moreover,
$$
  e^{\paralpha \pargX}P_0  h_{g,b}^{-1}P_0
 e^{-\paralpha \pargX}=
  \left \{e^{\paralpha \pargX}P_0 e^{-\paralpha \pargX}\right \}
  h_{g,b}^{-1}\left \{ e^{\paralpha \pargX}P_0
 e^{-\paralpha \pargX}\right \}
$$
is bounded due to \eqref{expP0}. Thus the only thing it remains to be
proved is the existence of a $b$ large enough such that the following
bound holds:
\begin{equation}\label{atreia10}
\sup_{g\in \bR}\left \Vert  e^{\paralpha \pargX} \left(1+\hat{B}_{g,b}\right)^{-1}
  e^{-\paralpha \pargX}\right \Vert <\infty.
\end{equation}
Using the Neumann series for  $\left( 1+\hat{B}_{g,b}\right)^{-1}$,
it follows that it suffices to prove that
\begin{equation}\label{B=0}
\lim_{b \rightarrow \infty}\sup_{g\in\bR}\left \Vert
 e^{\paralpha \pargX} \hat{B}_{g,b} e^{-\paralpha \pargX}\right \Vert=0.
\end{equation}
 Since (see \eqref{invh}) $\lim_{b \rightarrow \infty}\Vert
h_{g,b}^{-1}\Vert=0$
(uniformly in $g \in \bR$), for \eqref{B=0} to holds true it is
sufficient to show:
\begin{equation}\label{apatra11}
\sup_{g\in\bR}\left \Vert  e^{\paralpha \pargX}\left[P_{0},h_{g,b}\right] e^{-\paralpha
  \pargX}\right \Vert \leq {\rm const}.
\end{equation}
But this easily follows from \eqref{P0h}, \eqref{derh},
\eqref{prima11} and \eqref{adoua11} where we take $K=H$, $\alpha_z=\paralpha$
and $ h=g_g$. The proof of (iii) is concluded.

\vspace{0.5cm}

\noindent {\it Proof of} (iv). We start with a technical result:

\begin{lemma}\label{proiectorHS}
Fix $0\leq \peralpha<\sqrt{E_+}$. Then there exists a bounded operator $D$ such that
\begin{equation}\label{asasea11}
P_0= e^{-\peralpha \perX}(\bP^2 +1)^{-1}D
\end{equation}
\end{lemma}
\noindent{\it Proof}. We use the notation and ideas of Proposition \ref{locP0}, and
we rewrite $P_0$ in a convenient form.
First, for $R>0$ we have
$$
(H-z)^{-1}=
(H_R-z)^{-1}-(H_R-z)^{-1}\chi_R V(H-z)^{-1}.
$$
Second, choose $\Gamma$ close enough to $\sigma_0$ and $R$ large
enough, such that $(H_R-z)^{-1}$ becomes analytic
inside $\Gamma$ and \eqref{acincea6} holds true for all $z
\in \Gamma$. Then we can write:
\begin{align}\label{P01}
&P_0=-e^{-\peralpha \perX}\frac{i}{2\pi}\int_{\Gamma} \\
&(H_R-z)^{-1}[1+\peralpha (i\bP\cdot \nabla \pg +i \nabla \pg\cdot \bP
-\peralpha |\nabla \pg|^2)
(H_R-z)^{-1}]^{-1}e^{\peralpha \pg}\chi_R V(H-z)^{-1} dz.\nonumber
\end{align}
Now by the closed graph theorem we have that $(\bP^2+1)(H_R+1)^{-1}$ is
bounded (here $R$ is large enough such that $(-\infty, -1/2)\subset
\rho(H_R)$),
and together with the spectral
theorem:
$$\sup_{z\in\Gamma}\Vert (\bP^2+1)(H_R-z)^{-1}\Vert <\infty.$$
Use this in \eqref{P01} and we are done.
\qed

\vspace{0.5cm}

We now have all the necessary ingredients for proving the last
statement of our theorem. For every $L>0$ and $a\in \bR$, denote by
$\chi_{L,a}$ the characteristic function of the slab
$\{ \bx:\; |x_1 -a| \leq L \}$. Then define the operator
$B:=\chi_{L,a}P_0$. Using \eqref{asasea11} let us show that $B$ is
Hilbert-Schmidt, and moreover, uniformly in  $a\in
\bR$ we have:
\begin{equation}\label{prima100}
\Vert  B \Vert _2^2 \leq M\cdot L,
\end{equation}
for some $M< \infty$. Indeed, since $
B= \chi_{L,a}e^{-\peralpha \perX}(-\Delta
+1)^{-1}D$,
a direct computation using the explicit formula for the integral
kernel of the free Laplacian gives:
$$
\Vert \chi_{L,a}e^{-\peralpha \perX}(\bP^2
+1)^{-1}\Vert ^2_2 \leq {\rm const}\cdot L.
$$
It follows that the operator
$
 \chi_{L,a}P_0 \chi_{L,a}= BB^*
$
is trace class and
\begin{equation}\label{LaP}
\left \vert {\rm Tr} (\chi_{L,a}P_0\chi_{L,a})\right \vert \leq \Vert B\Vert_2^2\leq
M\cdot L
\end{equation}
for some $M< \infty$ independent of $L$ and $a$.

Now let $P_0^{L,a}$ be the orthogonal projection onto the subspace spanned
by those $W_{g,j}$ for which $g\in [a-L,a+L]$:
\begin{equation}\label{adoua100}
P_0^{L,a}:= \sum_{|g-a|\leq L}\sum_{j=1}^{m_g}\langle \cdot,
W_{g,j}\rangle W_{g,j}.
\end{equation}
We can choose $A$ sufficiently large such that \eqref{parW} implies:
\begin{equation}\label{apatra100}
\int _{|x_1-a| \geq A} | W_{g,j}(\bx )|^2 d\bx \leq \frac{1}{2},
\end{equation}
uniformly in $a$ and $g\in [a-L,a+L]$. Since $P_0 \geq P_0^{L,a}$, from  \eqref{LaP} one obtains:
\begin{align}\label{atreia100}
M\cdot (L+A)&\geq {\rm Tr} (\chi_{L+A,a} P_0 \chi_{L+A,a})\geq
{\rm Tr} (\chi_{L+A,a} P_0^{L,a} \chi_{L+A,a})\nonumber \\
 &=\sum_{|g-a|\leq L}\sum_{j=1}^{m_g}\int_{\bR^3}\chi_{L+A,a}(\bx)\vert
 W_{g,j}(\bx)\vert ^2 d\bx \nonumber \\
&\geq
\sum_{|g-a|\leq L}\sum_{j=1}^{m_g}\frac{1}{2}=\frac{1}{2}N(a, L),
\end{align}
where in the last inequality we used \eqref{apatra100}.
In particular, if $L \geq 1$, then uniformly in $a\in
\bR$ we have
$$
N(a, L) \leq 2M\cdot (1+A)L
$$
and the proof is finished.

\subsection{Proof of Proposition \ref{optimal}}

The only thing we have to prove is that (\ref{apatra11}) holds
true for $\paralpha$ replaced by any $\alpha<\alpha_0$, where
$\alpha_0$ is the a-priori given, "exact" exponential localization. 

We introduce the multiplication operator given by 
$\{e^{\alpha|\cdot-t|}f\}(\bx):=e^{\alpha|x_1-t|}f(\bx)$. 
We start by noticing that due to the bound $
e^{\pm \alpha(\sqrt{s^2 +1}-\vert s \vert)} \leq e^{\alpha}$ we can
replace \eqref{decP} with:
\begin{equation}\label{decPP}
 \sup_{t\in \bR}\Vert e^{\alpha|\cdot-t|} P_0 e^{-\alpha |\cdot -
   t|}\Vert <\infty.
\end{equation}
The same replacement can be done in \eqref{apatra11}. Now the integral
kernel $\cA (\bx,\by)$ of the operator $ A:=e^{\alpha |\cdot -g|}\left[P_{0},h_{g,b}\right]
 e^{-\alpha |\cdot-g|}$ equals
  \begin{equation}\label{ultima10}
\cA (\bx,\by)=
  \cP_0(\bx,\by)e^{\alpha(|x_1-g| -|y_1-g|)}(h_{g,b}(\by)-h_{g,b}(\bx)).
\end{equation}
We consider $A$ as an operator on $L^2(\bR^3)=\bigoplus_{p\in
  \bZ}L^2([p,p+1]\times \bR^2)$. Let $\chi_p$ be the characteristic
function of the slab $[p,p+1]\times \bR^2$. We have that $A_{pp'}:=\chi_pA\chi_{p'}$ is a
bounded operator between $L^2([p',p'+1]\times \bR^2)$ and $L^2([p,p+1]\times \bR^2)$, and we can write 
$A=\{A_{pp'}\}_{p,p'\in \bZ}$.  We will bound the norm of $A$ with a 
Schur-Holmgren type estimate (see below Lemma \ref{schurholmgren}):
\begin{equation}\label{margy222}
||A||\leq 
\left (\sup_{p'\in \bZ}\sum_{p\in\bZ}||A_{pp'}||\right )^{\frac{1}{2}} \left (\sup_{p\in
    \bZ}\sum_{p'\in\bZ}||A_{pp'}||\right )^{\frac{1}{2}} .
\end{equation} 
For $0\leq x_1,y_1\leq 1$, the kernel of $A_{pp'}$ can be written as:
\begin{align}\label{margy223}
\cA_{pp'} (\bx,\by)&=\cP_0(x_1+p,\px;y_1+p',\py)e^{\alpha(|x_1+p-g|
  -|y_1+p'-g|)}
(h_{g,b}(y_1+p')-h_{g,b}(x_1+p))\nonumber \\
&=\cP_0(x_1+p,\px;y_1+p',\py)e^{\alpha(|x_1+p-g|
  -|y_1+p'-g|)}
(h_{g,b}(p')-h_{g,b}(p))\nonumber \\
&+\cP_0(x_1+p,\px;y_1+p',\py)e^{\alpha(|x_1+p-g|
  -|y_1+p'-g|)}
(h_{g,b}(y_1+p')-h_{g,b}(p'))\nonumber \\
&+\cP_0(x_1+p,\px;y_1+p',\py)e^{\alpha(|x_1+p-g|
  -|y_1+p'-g|)}
(-h_{g,b}(x_1+p)+h_{g,b}(p))\nonumber\\
 &=:\cA_{pp'}^{(1)} (\bx,\by)+\cA_{pp'}^{(2)} (\bx,\by)+\cA_{pp'}^{(3)} (\bx,\by).
\end{align}
The last two kernels can be analyzed with the same methods as the
first one, thus we only estimate the norm of $A_{pp'}^{(1)}$. The
crucial observation is that we can write this operator as a product of
three operators having the corresponding kernels:
\begin{align}\label{margy224}
&\cA_{pp'}^{(1)} (\bx,\by)=e^{\alpha(|x_1+p-g|-|p-g|)}\nonumber \\
&\cdot e^{\alpha(|p-g|
  -|p'-g|)} \cP_0(x_1+p,\px;y_1+p',\py)(h_{g,b}(p')-h_{g,b}(p))\nonumber \\
&\cdot e^{-\alpha(|y_1+p'-g|-|p'-g|)}.
\end{align}
The kernel in the middle corresponds to the operator
$\chi_pP_0\chi_p'$ times some coefficients depending on $p,p'$.  

Using the triangle inequality to bound the exponentials, and
(\ref{derh}) in order to write $
\vert h_{g,b}(\by)-h_{g,b}(\bx)\vert \leq K \vert x_1-y_1 \vert
$, we have:
$$ ||A_{pp'}^{(1)}||\leq K e^{2\alpha} e^{\alpha |p-p'|}|p-p'|\cdot 
||\chi_pP_0\chi_p'||.$$
Using $t=p'$ and $(\alpha+\alpha_0)/2$ in \eqref{decPP} we obtain
$$||\chi_pP_0\chi_p'||\leq C e^{-(\alpha+\alpha_0)|p-p'|/2},$$
thus 
$$||A_{pp'}^{(1)}||\leq C' |p-p'|e^{-(\alpha_0-\alpha)|p-p'|/2}$$
which is summable in the sense of \eqref{margy222}. 
The same strategy can be applied in the case of  $A_{pp'}^{(2)}$ and
$A_{pp'}^{(3)}$. The last thing to be done is to prove the
Schur-Holmgren estimate:
\begin{lemma}\label{schurholmgren}
The estimate \eqref{margy222} holds true.
\end{lemma}  
\begin{proof}
Let $\psi\in L^2(\bR^3)$ with compact support and $||\psi||=1$. We write:
\begin{align}\label{kkjja}
||A\psi||^2=\sum_{p\in \bZ}||\chi_pA\psi||^2.
\end{align}
But
\begin{align}\label{kkjja2}
||\chi_pA\psi||&\leq \sum_{p'\in
  \bZ}\sqrt{||A_{pp'}||}\;\sqrt{||A_{pp'}||} \;||\chi_{p'}\psi||\leq  \left \{ \sum_{p'\in
  \bZ}||A_{pp'}||\right\}^{\frac{1}{2}} \left \{ \sum_{p'\in
  \bZ}||A_{pp'}||\; ||\chi_{p'}\psi||^2\right\}^{\frac{1}{2}}\nonumber
\\
&\leq \left \{ \sup_{s\in \bZ}\sum_{t\in
  \bZ}||A_{st}||\;\right\}^{\frac{1}{2}}\left \{ \sum_{p'\in
  \bZ}||A_{pp'}||\; ||\chi_{p'}\psi||^2\right\}^{\frac{1}{2}}
\end{align}
where in the second inequality we used Cauchy-Schwarz with respect to
$p'$. Introduce this in \eqref{kkjja} and the bound follows after the
use of $\sum_{p'\in
  \bZ}||\chi_{p'}\psi||^2=1$.
\end{proof}

\vspace{1cm}

\noindent{\bf Acknowledgements.} Part of this work was done during a visit of G. Nenciu
at the Department of Mathematical Sciences, Aalborg University;
 both hospitality and financial support are gratefully acknowledged.
H. Cornean acknowledges support from Danish F.N.U. grant
{\it Mathematical Physics and Partial Differential Equations}.
A. Nenciu and G. Nenciu  were partially supported by CEEX Grant 05-D11-45/2005.
We also thank the first referee for his/hers most valuable comments regarding Proposition \ref{optimal}.

\end{document}